\documentclass[pre,aps,showpacs,twocolumn]{revtex4}
\usepackage{amsmath}
\usepackage{amssymb}
\usepackage{bm}
\usepackage[dvips]{graphicx}

\begin{document}

\title{Density nonlinearities in field theories for a toy model of fluctuating nonlinear hydrodynamics of supercooled liquids}
\author{Joonhyun Yeo}\email{jhyeo@konkuk.ac.kr}
%

\affiliation{Division of Quantum Phases and Devices, School of Physics, Konkuk University, Seoul 143-701, Korea}

\date{\today}
\begin{abstract} 
We study a zero-dimensional version of the fluctuating nonlinear
hydrodynamics (FNH) of supercooled liquids originally investigated by 
Das and Mazenko (DM) [Phys.\ Rev.\ A {\bf 34}, 2265 (1986)].
The time-dependent density-like and momentum-like variables
are introduced with no spatial degrees of freedom in this toy model. 
The structure of nonlinearities takes the similar form to the 
original FNH, which allows one to study in a simpler setting 
the issues raised recently regarding the field theoretical 
approaches to glass forming liquids.
We study the effects of density nonlinearities on the time evolution of correlation
and response functions by developing field theoretic formulations in two different ways:
first by following the original prescription of DM and then by constructing a dynamical 
action which possesses a linear time reversal symmetry as proposed recently. 
We show explicitly that, at the one-loop order of the perturbation theory, 
the DM-type field theory does not support a sharp
ergodic-nonergodic transition, while the other admits one.
The simple nature of the toy model in the DM formulation 
allows us to develop numerical solutions to 
a complete set of coupled dynamical equations for the correlation and response functions
at the one-loop order. 
\end{abstract}




\maketitle
\section{Introduction}
The slow dynamics of supercooled liquids near the glass transition has been under
intense theoretical and experimental investigation for many years. 
Among many theoretical attempts to understand the slowing down of supercooled liquids,
the mode coupling theory (MCT) \cite{Gotze0,KM,Das,RC} stands out
as one of the most successful ones. It explains, for example,
an elaborate sequence of time relaxation processes with characteristic exponents
which are consistent with experimental findings. 
In its initial form \cite{Leuth,Bengtz,DMRT}, referred to as the standard MCT, 
it predicts a sharp ergodic-to-nonergodic (ENE)
transition at a critical temperature or density with the nonergodic phase
characterized by the density autocorrelation function approaching 
a nonzero value called the nonergodicity parameter
in the long time limit. 
Many experiments and numerical simulations, however, show that this
feature is not realized in real supercooled liquids in finite dimensions,
and that the ergodicity is restored at finite temperature. 

There have been many attempts to put the MCT into
a field theoretic framework \cite{DM,SDD,MR,ABL,BBB,NH,KaM,KK,Sz,BR}, 
since it has many advantages including the possibility 
of a systematic improvement. Das and Mazenko (DM) \cite{DM}
studied the nonlinear feedback mechanism of density fluctuations in supercooled liquids
by formulating a field theoretic renormalized perturbation theory
of the fluctuating nonlinear hydrodynamics (FNH) of compressible fluids.
They find that the sharp transition is cutoff and the system remains ergodic at all temperatures or densities.
Recently, however, the validity of the DM results, especially of those on the explanation
of the cutoff mechanism, was questioned in Refs.~\cite{ABL,CR,NH}.
Some of these works are based on a field theory developed in
Ref.~\cite{ABL} where the dynamical action is invariant
under a set of linear time-reversal transformations. This formulation allows one 
to have a full set of fluctuation-dissipation
relations (FDR) relating linearly correlation functions to response functions. 
The field theory of DM has only a limited number of linear FDR and some relations
hold only in the hydrodynamic limit. This is one of the points on which 
the conclusion by DM on the absence of the sharp ENE transition was questioned. 
The field theory with linear FDR was later improved \cite{KK} for the case of interacting Brownian particles 
satisfying the Dean-Kawasaki equation \cite{Dean,Kawasaki},
where the standard MCT result was recovered at the one-loop order of perturbation theory. 
This improved method was then applied to the FNH \cite{NH} with results indicating 
a sharp ENE transition at the one-loop order with 
the nonergodicity parameter satisfying the standard MCT result.
In response to these developments, DM reexamined their work and showed \cite{DM2} in 
a nonperturbative analysis without resorting to the hydrodynamic limit that
the sharp ENE transition is not present in the FNH after all.  
This conclusion is also supported by the recent direct numerical integration
of the generalized Langevin equations of the FNH \cite{Gupta}.

It is somewhat puzzling to have completely different results from the two field theoretic approaches 
of the same model. In this respect, it might be desirable to have a simpler setting
in which one can compare these two field theoretical approaches
and study where the difference originates.
The field theoretical treatment of FNH is complicated by many factors including the presence of 
many dynamical variables. This is especially the case
for the field theory with linear FDR. 
In this paper, we introduce a simple toy model of FNH,
which can shed some light on the issues described above concerning the field theoretic approaches to the FNH.
In this toy model, there is no spatial dependence in the
dynamical field variables which consist simply of a density-like variable
and a single component momentum-like variable. We develop the two different field theories
of the toy model, namely the original
DM-type field theory and the one with linear FDR.
We show explicitly that, at the one-loop order of perturbation expansion, 
a sharp ENE-type transition does not occur in 
the DM-type field theory. On the other hand, the field theory with linear FDR
shows a sharp transition at the one-loop order. 
By comparing the two field theories, we find that
the major difference between the two formulations
lies in the way of treating the density nonlinearities present in FNH
within the renormalized perturbation theory.
In particular, the field theory with linear FDR results in a dynamical action
which contains non-polynomial functions of field variables in contrast to the DM
field theory. This implies that,
when the renormalized perturbation theory is performed at a given order of 
the loop expansion, the two field theories end up with treating the density nonlinearities in a
different way, since the field theory with linear FDR involves truncating the non-polynomial functions.

Although there is only a limited number of linear FDR,
the DM field theory at a given order of the loop expansion 
can be regarded as a well-defined self-consistent theory among the correlation and response functions 
satisfying a set of coupled self-consistent equations.
In this paper, we construct the set of coupled equations at the one-loop order
for the DM field theoretic approach to the toy model and study them numerically.
Since the simple nature of the toy model reduces the number of independent
correlation and response functions, we were able to solve these
equations numerically. 

In the next section, we introduce our toy model and construct the field theories following
the DM prescription and the method involving linear FDR, respectively. In Sec.~\ref{sec:oneloop},
we study the time evolution of the correlation functions using the Schwinger-Dyson equations
for both field theoretic formulations. We then analyze the time evolution equations 
for the correlation function corresponding to the density autocorrelation function 
of the FNH at the one-loop order for 
the possible existence of a sharp ENE-type transition. 
In Sec.~\ref{sec:numerical}, we present a set of
coupled equations for the correlation and response functions in the DM field theory
at the one-loop order and their numerical solutions.
In the final section, we summarize our results
with discussion.

\section{Model}

Our model is a zero-dimensional version of the FNH of compressible fluids 
developed by Das and Mazenko \cite{DM}.
We introduce as our dynamical variables a time-dependent density-like variable $a(t)$
and a single-component momentum-like variable $b(t)$ without any spatial 
degrees of freedom. 
In order to construct the equations motion for these variables, we introduce the
effective free energy $F$ where the equilibrium distribution for the system at 
temperature $T$ is given by $\exp(-F/T)$. The free energy can be written as
$F=F_K+F_U$ where $F_K[a,b]$ is
the kinetic energy and the potential energy part $F_U[a]$ is assumed to 
depend only on $a$. We take the usual form for the kinetic part, that is
$
F_K=\frac{b^2}{2a}.
$
The nonlinearity in the form of $1/a$ plays an important role in the following discussion.

Kim and Kawasaki \cite{KKtoy} introduced a similar zero-dimensional toy model
of the FNH to the present one by incorporating the multicomponent density-like and momentum-like variables. 
However, in addition to having multicomponent fields, their model differs from ours in a 
fundamental way. Their free energy is quadratic both in the density-like and the momentum-like
variables without the $1/a$ nonlinearity which is present in the FNH of compressible fluids.
In the present toy model, therefore,
the actual form of the equations of motion will be different from those in Ref.~\cite{KKtoy}, 
but the derivation of the equations from the free energy can be performed in the same way.  
In our model, the equation of motion for the variable $a(t)$ takes 
the form of a zero-dimensional version of the 
continuity equation, namely
\begin{equation}
\dot{a}(t)+J b(t)=0
\label{eq:a}
\end{equation}
for some constant $J$. This can be regarded as the reversible dynamics for $a(t)$ which can be
derived from  
\begin{equation}
\dot{a}(t)=Q_{ab} \frac{\partial F}{\partial b}-T\frac{\partial Q_{ab}}{\partial b},
\end{equation}
with $Q_{ab}=-Ja$ playing the role of the Poisson bracket. 
The equation of motion for $b(t)$ has the dissipative part described by the coefficient $\Gamma$ 
in addition to the reversible part as follows:  
\begin{equation}
\dot{b}(t) = Q_{ba}\frac{\partial F}{\partial a} -T\frac{\partial Q_{ba}}{\partial a}
-\Gamma \frac{\partial F}{\partial b} +\theta(t), 
\end{equation}
where $Q_{ba}=-Q_{ab}$ and 
the Gaussian white noise $\theta$ has the variance 
$\langle\theta(t)\theta(t^\prime)\rangle=2\Gamma T\delta(t-t^\prime)$.
In the present work, we take the simple quadratic form for 
the potential energy part $F_U$ of
the effective free energy $F$ as
\begin{equation}
F_U[a]=\frac{A}{2}(\delta a)^2
\end{equation}
with the fluctuation $\delta a = a- a_0$ and the average value $a_0$.
We thus have the equation of motion for $b(t)$ as
\begin{equation}
\dot{b}(t) +J\left(\frac{b^2}{2a}\right)-JAa(\delta a) + TJ
+\Gamma \left(\frac{b}{a}\right) =\theta, \label{eq:b} 
\end{equation}

We can easily verify that the equilibrium stationary distribution corresponding to 
the above equations for $a$ and $b$ is proportional to $\exp(-F/T)$.
The corresponding Fokker-Planck equation for the probability 
distribution $P(a,b,t)$ is given by $\partial_t P = {\cal L}P$ where
the Fokker-Planck operator is given by
${\cal L}={\cal L}_1 + {\cal L}_2$ with
\begin{eqnarray}
&&{\cal L}_1=\frac{\partial}{\partial b} \Gamma \left( T\frac{\partial}{\partial b} + \frac{b}{a}\right) \\
&&{\cal L}_2=\frac{\partial}{\partial a} (Jb) +\frac{\partial}{\partial b}
\left( J\left(\frac{b^2}{2a}\right)-JAa(\delta a) + TJ \right).
\end{eqnarray}
It is straightforward to show that $P\sim\exp(-F/T)$ satisfies
${\cal L}P=0$.

One can develop a field theory from the above Langevin equations by using the standard
Martin-Siggia-Rose (MSR) formalism \cite{MSR}. In the MSR procedure, the hatted fields
$\hat{a}$ and $\hat{b}$ are introduced to enforce the equations of motion for $a$ and $b$ respectively.  
We present in the following subsections two different field theoretical approaches to this model
and compare the outcomes of both approaches concerning the existence of an ENE transition. 
The first one is the original Das and Mazenko approach \cite{DM}, where the $1/a$ nonlinerarities
in the model are taken care of in a simple way by the introduction of a single additional auxiliary field. 
The second approach due to Refs.~\cite{ABL,KK,NH} incorporates the linear time-reversal 
symmetry into the dynamical action resulting in a set of linear FDR. In order to do that more auxiliary 
fields will have to be introduced.

\subsection{Das-Mazenko approach}

In the DM approach, an auxiliary velocity-like field $c(t)$ is introduced
such that the condition $b(t)=a(t)c(t)$ is enforced through a delta function
\begin{eqnarray}
1&=&\int{\cal D}c(t)\;\delta\left(a(t)c(t)-b(t)\right) \nonumber \\
&=&\int{\cal D}c(t) \int {\cal D} \hat{c}(t) 
\exp\left[-i \hat{c}(t) \left(b(t)-a(t)c(t)\right) \right] .
\end{eqnarray} 
The first equality holds up to a Jacobian.
This Jacobian was shown in Ref.~\cite{MazYeo} to have no effect on the correlation and response functions
and will be neglected in the following analysis. 
Using this identity, we obtain the generating functional $Z$ as a functional integral over
the fields $\psi_i(t)=\delta a(t), b(t), c(t)$ and $\hat{\psi}_i (t)=\hat{a}(t), \hat{b}(t), \hat{c}(t)$. 
We can write
$Z_{\rm DM}=\int\prod_i {\cal D}\psi_i {\cal D} \hat{\psi}_i \; \exp(-S_{\rm DM}[\psi,\hat{\psi}])$, where
\begin{eqnarray}
&&S_{\rm DM}=\int dt \;\; [ \Gamma T \hat{b}^2(t) 
    +i \hat{a}(t)\left\{ \dot{a}(t)+Jb(t)\right\}  \nonumber \\
&&\quad\quad\quad +i\hat{b}(t)\{ \dot{b}(t) -JAa_0 \delta a(t) + \Gamma c(t) +TJ \nonumber \\
&&\quad\quad\quad +\frac{Ja_0}{2} (c(t))^2 + \frac{J}{2} 
                           \delta a(t) (c(t))^2 -JA (\delta a(t))^2\} \nonumber \\
&&\quad\quad\quad +i\hat{c}(t)\{ b(t) -a_0 c(t) -\delta a(t)c(t) \} ] .
\label{S_DM}
\end{eqnarray}

We will use $\Psi(t)$ to represent any one of the six variables $\{\psi_i,\hat{\psi_i}\}$ in our model, and 
denote the two-point
correlation function between arbitrary two variables $\Psi(t)$ and $\Psi^\prime(t^\prime)$ by
\begin{equation}
 G_{\Psi\Psi^\prime}(t-t^\prime)=\langle \Psi(t) \Psi^\prime(t^\prime) \rangle.
\end{equation}
(For the subscripts of $G$, we will use $a$ instead of $\delta a$ for simplicity.) 
Note that among the correlation functions those between two hatted variables vanish due to causality, that is
$G_{\hat{\psi}_i\hat{\psi}_j}=0$. It follows that $iTJ\hat{b}(t)$ term in the action Eq.~(\ref{S_DM})
has no effect on the correlation functions and will be neglected in the following.
The causality also requires that $G_{\psi_i\hat{\psi}_j}(t)=0$ for $t<0$.

We can easily establish some nonperturbative relations among the correlation functions which will be useful in later discussion. 
If we use $\Psi(t^\prime)=\psi_i(t^\prime)=\delta a(t^\prime),b(t^\prime)$ and $c(t^\prime)$
in the identity
\begin{equation}
 0=\int\prod_i {\cal D}\psi_i {\cal D} \hat{\psi}_i \; \frac{\delta}{\delta \hat{a}(t)}\left[\Psi(t^\prime)
\exp(-S_{\rm DM})\right],
\end{equation}
we obtain
\begin{equation}
\dot{G}_{a\psi}(t)+JG_{b\psi}(t)=0.
\label{np1}
\end{equation}
On the other hand, if $\Psi(t^\prime)=\hat\psi(t^\prime)=\hat{a}(t^\prime),\hat{b}(t^\prime)$ 
and $\hat{c}(t^\prime)$ are used, we have
\begin{equation}
\dot{G}_{a\hat{\psi}}(t)+JG_{b\hat{\psi}}(t)
=-i\delta_{\hat{\psi}\hat{a}}\delta(t).
\label{np2}
\end{equation}
Note that Eqs.~(\ref{np1}) and (\ref{np2}) are direct consequences of the zero-dimensional
version of the mass conservation law given by Eq.~(\ref{eq:a}).
In the DM approach, only a limited number of FDR exist that relate linearly the correlation functions 
to response functions.
Assuming the time reversal properties of the fields as $a(-t)=a(t)$, $b(-t)=-b(t)$
and $c(-t)=-c(t)$, we can derive the FDR for $\psi=a,b$ and $c$ as
\begin{equation}
G_{\psi\hat{b}}(t)=-\frac{i}{T} \Theta(t)G_{\psi c}(t),
\label{fdr1}
\end{equation}
where $\Theta(t)=1$ for $t>0$ and vanishes for $t<0$.
The detailed derivation of the FDR closely follows the one given in Ref.~\cite{DM}.
Since $\hat{\psi}_i$ is a real field,
we can show that the correlation function between unhatted and hatted variables is a pure imaginary
number, that is
\begin{equation}
G_{\psi_i\hat{\psi}_j}^\ast(t)=-G_{\psi_i\hat{\psi}_j}(t).
\end{equation}

\subsection{Field theory with linear FDR}

We apply the field theoretical approach developed in Refs.~\cite{ABL,KK,NH}
to our toy model given by the dynamic equations (\ref{eq:a}) and (\ref{eq:b}). 
Among these methods, we will follow closely the improved procedure described in Ref.~\cite{NH}.
In order to do that, we introduce two
auxiliary fields $\eta(t)$ and $\theta (t)$ defined by
\begin{eqnarray}
\eta &=& \frac{\partial F}{\partial b} - \frac{b}{a_0} 
=-\frac{b}{a_0}\sum_{k=1}^{\infty}(-1)^{k}\left(\frac{\delta a}{a_0}\right)^k \nonumber \\
&\equiv& f_\eta(\delta a,b) 
\label{eta_def} \\
\theta &=& \frac{\partial F}{\partial a} - A (\delta a) = -\frac{b^2}{2a_0^2}\sum_{k=0}^{\infty}
(-1)^k \left(\frac{\delta a}{a_0}\right)^k \nonumber \\
&\equiv& f_\theta (\delta a,b).
\label{theta_def}
\end{eqnarray}
Note that these are non-polynomial functions of the main dynamical variables $\delta a$ and $b$. 
Introducing the hatted counterparts $\hat\eta(t)$ and $\hat\theta(t)$ 
to enforce these definitions for the new variables, we can construct
the generating functional as $Z_{\rm FDR}=\int\prod_i {\cal D}\phi_i {\cal D}\hat{\phi}_i \; 
\exp(-S_{\rm FDR}[\phi,\hat{\phi}])$, where $\phi_i (t)=\delta a(t), b(t), \eta(t), \theta(t)$ and
$\hat{\phi}_i (t)=\hat{a}(t), \hat{b}(t), \hat{\eta}(t), \hat{\theta}(t)$. 
Similarly to the DM case, we will use $\Phi(t)$ to represent one of the eight variables $\{\phi_i,\hat{\phi_i}\}$. 
The dynamical action 
in this case can be written as a sum of the Gaussian and the nonlinear parts as
$S_{\rm FDR}=S^{(0)}_{\rm FDR}+S^{(1)}_{\rm FDR}$, where 
\begin{eqnarray}
&& S^{(0)}_{\rm FDR} = \int dt \;\; [ \Gamma T \hat{b}^2(t) 
    +i \hat{a}(t)\{ \dot{a}(t)+Jb(t) +J a_0 \eta(t) \}  \nonumber \\
&&\quad +i\hat{b}(t)\{ \dot{b}(t) -JAa_0 \delta a(t) -Ja_0 \theta (t) 
+ \Gamma c(t) +\frac{\Gamma}{a_0} b(t)\}  \nonumber \\
&&\quad +i\hat{\eta}(t)\eta(t) +i \hat{\theta} (t) \theta(t) ] ,
\end{eqnarray}
and
\begin{eqnarray}
S^{(1)}_{\rm FDR}&=&
\int dt \;\; [ i \hat{a}(t)\{ J\delta a(t) \eta(t) +\frac{J}{a_0} \delta a(t) b(t) \}  \nonumber \\
&&\quad\quad -i\hat{b}(t)\{ J \delta a(t) \theta (t) +JA (\delta a (t))^2 \}  \nonumber \\
&&\quad\quad -i\hat{\eta}(t) f_\eta (\delta a,b) 
-i\hat{\theta}(t)f_\theta (\delta a,b) ] .
\end{eqnarray}
Note that we have used the identity 
\begin{equation}
a_0 c(t)+ \delta a(t)\eta(t) + \frac 1 {a_0} \delta a(t) b(t) =0,
\label{id}
\end{equation}
which follows directly from the definition of $\eta(t)$, Eq.(\ref{eta_def}).
The above actions $S^{(0)}_{\rm FDR}$ and
$S^{(1)}_{\rm FDR}$ are separately invariant under the time reversal transformations given by
\begin{eqnarray}
&& \delta a(-t)=\delta a(t), \quad b(-t)=-b(t), \nonumber \\
&& \eta(-t)=-\eta(t), \quad \theta (-t)=\theta (t), \nonumber \\
&&\hat{a}(-t)= -\hat{a}(t) -\frac{i}{T}\theta (t) -i\frac{A}{T} \delta a(t), \nonumber \\ 
&&\hat{b}(-t)=\hat{b}(t) + \frac{i}{T} \eta(t)+\frac{i}{a_0 T} b(t), \nonumber \\
&&\hat{\eta}(-t)= -\hat{\eta}(t)+\frac{i}{T}\dot{b}(t), \nonumber \\
&&\hat{\theta}(-t)=\hat{\theta}(t)-\frac{i}{T}\dot{a}(t) .
\label{TRS}
\end{eqnarray}
Applying the time reversal invariance on the various correlation functions 
by following the procedures described in Refs.~\cite{ABL,KK,NH},
we obtain a set of FDR. Here we only list those which are relevant to 
the discussion in the next section. We have for $\phi=a,b,\eta$ and $\theta$
\begin{eqnarray}
&&G_{\phi \hat{a}}(t)=-\frac{i}{T}\Theta(t)[AG_{\phi a}(t)+G_{\phi\theta}(t)], \nonumber \\
&&G_{\phi \hat{b}}(t)=-\frac{i}{T}\Theta(t)[\frac{1}{a_0}G_{\phi b}(t)+G_{\phi\eta}(t)], \nonumber \\
&&G_{\phi \hat{\eta}}(t)=-\frac{i}{T}\Theta(t)\dot{G}_{\phi b}(t), \nonumber \\
&&G_{\phi \hat{\theta}}(t)=-\frac{i}{T}\Theta(t)\dot{G}_{\phi a}(t) .
\label{fdr_fdr_1}
\end{eqnarray}

\section{Renormalized Perturbation Theory: One-Loop Order}
\label{sec:oneloop}

\begin{figure}
 \begin{center}
\includegraphics[width=0.4\textwidth]{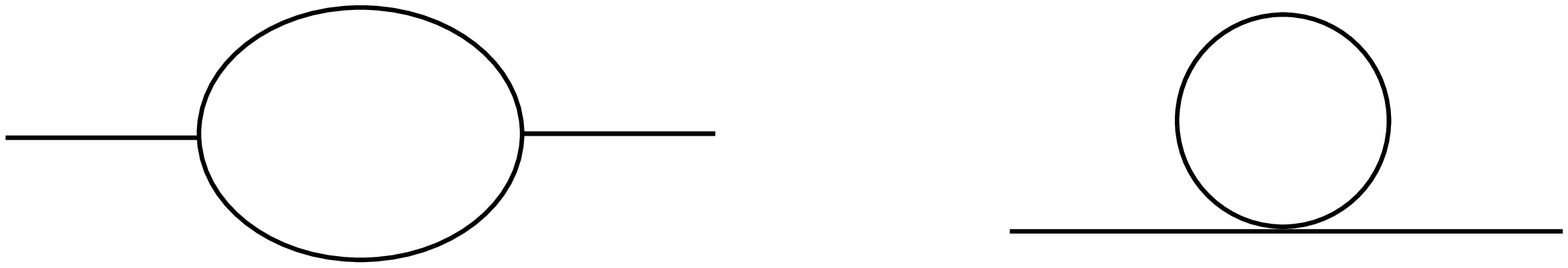}
\end{center}
\caption{The one-loop diagrams contributing to the self-energy $\mathsf{\Sigma}$.}
\label{diagram_ab}
\end{figure}

In this section, we develop self-consistent renormalized perturbation theories for the two field theoretic approaches
introduced in the previous section. We then focus on the time evolution of two-point correlation functions
using the Schwinger-Dyson (SD) equation. In particular, we study the $t\to\infty$ limit of $G_{aa}(t)$, which
corresponds to the density auto-correlation function in FNH, to explore the possibility of 
an ENE transition. The formal development of the self-consistent perturbation theory 
can be found in Refs.~\cite{ft1,ft2,KK}. The SD equation defines the self-energy $\mathsf{\Sigma}$ 
through its relation to the propagator (the two-point correlation function) $\mathsf{G}$.
It is given symbolically by
\begin{equation}
\mathsf{G^{-1}=G^{-1}_0-\Sigma},
\end{equation}
where the subscript $0$ refers to the bare quantity obtained by keeping only the Gaussian terms in the action.
The self-energy is obtained by differentiating the so-called two-particle irreducible vertex function 
$\Gamma_{\rm 2PI}[\mathsf{G}]$ with respect to the propagator $\mathsf{G}$.
At the one-loop order of the loop expansion of $\Gamma_{\rm 2PI}$, there are only two kinds of diagrams
for the self-energy
which are relevant to the two field theoretical approaches studied in the previous section. 
These are shown in Fig.~\ref{diagram_ab}.

\subsection{Das-Mazenko approach}

The SD equation for the DM field theory
between arbitrary two fields $\Psi$ and $\Psi^\prime$ is given by 
\begin{eqnarray}
&&\delta_{\Psi \Psi^\prime}\delta(t-t^\prime) \nonumber \\
&&=\sum_{\Psi^{\prime\prime}}\int dt^{\prime\prime} \{
\left[ G_0^{-1} \right]_{\Psi \Psi^{\prime\prime}}(t-t^{\prime\prime}) 
G_{\Psi^{\prime\prime} \Psi^\prime}(t^{\prime\prime}-t^\prime) \nonumber\\
&&  \quad\quad \quad\quad\quad\quad -\Sigma_{\Psi \Psi^{\prime\prime}} (t-t^{\prime\prime})
G_{\Psi^{\prime\prime} \Psi^\prime}(t^{\prime\prime}-t^\prime) \}.
\label{sd_general}
\end{eqnarray}
We note that the causality requirement for the self-energy reads $\Sigma_{\hat{\psi}_i\psi_j}(t)=0$
for $t<0$ and $\Sigma_{\psi_i \psi_j}(t)=0$ for all $t$. Similarly to the propagators, 
the self-energy between unhatted and hatted variables is pure imaginary.
Since there is no nonlinear term containing $b(t)$ in Eq.~(\ref{S_DM}), the self-energies involving $b$
must vanish. From this feature of the DM field theory, we can derive nonperturbative 
relations among correlation functions as
\begin{eqnarray}
&&JG_{a\hat{a}}(t)+\dot{G}_{a\hat{b}}(t)+G_{a\hat{c}}(t)=0, 
\label{abc1}\\
&&JG_{c\hat{a}}(t)+\dot{G}_{c\hat{b}}(t)+G_{c\hat{c}}(t)=0
\label{abc2}
\end{eqnarray}
from the $(\Psi,\Psi^\prime)=(b,a)$ and $(b,c)$ component of the SD equation, respectively.

\begin{figure}
 \begin{center}
\includegraphics[width=0.3\textwidth]{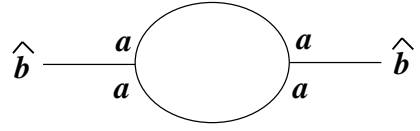}
\end{center}
\caption{The one-loop diagram for $\Sigma_{\hat{b}\hat{b}}$ that may have a non-vanishing
contribution in the $t\to\infty$ limit.}
\label{diagram}
\end{figure}

In order to study the time evolution of $G_{aa}(t)$, we look at
the $(\Psi,\Psi^\prime)=(\hat{b},a)$ component, which yields
\begin{eqnarray}
&&\dot{G}_{ba}(t)+\Gamma G_{c a}(t)-JAa_0 G_{aa}(t)-2i\Gamma T G_{\hat{b}a}(t)
\nonumber \\
&&=F_{\hat{b}a}(t),
\end{eqnarray}
where 
\begin{eqnarray}
&&F_{\hat{b}a}(t) \label{sd_bha} \\
&&= -i\int_{-\infty}^t ds\; \left\{ \Sigma_{\hat{b}a}(t-s)G_{aa}(s)
+\Sigma_{\hat{b}c}(t-s)G_{ca}(s) \right\} \nonumber\\
&& \quad
-i\int_{-\infty}^0 ds\; \left\{ \Sigma_{\hat{b}\hat{b}}(t-s)G_{\hat{b}a}(s)
+\Sigma_{\hat{b}\hat{c}}(t-s)G_{\hat{c}a}(s) \right\} .
\nonumber
\end{eqnarray}
Using Eqs.~(\ref{np1}) and (\ref{fdr1}), we can rewrite this equation for $t>0$ as
\begin{equation}
\ddot{G}_{aa}(t)-J\Gamma G_{c a}(t)+J^2Aa_0 G_{aa}(t)
=-JF_{\hat{b}a}(t),
\label{sd_bha:2}
\end{equation}
Now in order to investigate the possible ENE
transition in this model, we consider the $t\to\infty$ limit. 
Let us assume that all the other correlation
functions except $G_{aa}(t)$ vanish in the $t\to\infty$ limit. 
To the one-loop order of the perturbation expansion, only the self-energy $\Sigma_{\hat{b}\hat{b}}(t)$
can be nonvanishing in the $t\to\infty$ limit due to the diagram shown in 
Fig.~\ref{diagram}, since it is proportional to $(G_{aa}(t))^2$. 
Therefore, the nonvanishing contributions to 
$F_{\hat{b}a}(\infty)$ come from the first and the third terms on the right hand side of 
Eq.~(\ref{sd_bha}).
Among the terms on the left hand side of Eq.~(\ref{sd_bha:2})
only the third term is nonvanishing in this limit. Using Eq.~(\ref{fdr1}), we therefore have
\begin{equation}
JAa_0 G_{aa}(\infty) =\sigma G_{aa}(\infty) 
        + \frac {1}{T}\Sigma_{\hat{b}\hat{b}}(\infty)\int_0^\infty ds\; G_{ac}(s),
\label{gaa_infty}
\end{equation}
where $\sigma=\int_0^\infty ds\; i\Sigma_{\hat{b}a}(s)$ is a finite real number.
The relation between $G_{ac}(t)$ and $G_{aa}(t)$ can be obtained from the 
$(\Psi,\Psi^\prime)=(\hat{c},a)$ component 
of the SD equation, which is given by
\begin{equation}
-\frac{1}{J}\dot{G}_{a a}(t)
+a_0 G_{ac}(t) =F_{\hat{c}a}(t),
\label{gaa_gac}
\end{equation}
where
\begin{eqnarray} 
&& F_{\hat{c}a}(t) \label{sd_cha} \\
&&= -i\int_{-\infty}^{t} ds\; \left\{ \Sigma_{\hat{c}a}(t-s)G_{aa}(s)
+\Sigma_{\hat{c}c}(t-s)G_{ca}(s) \right\} 
\nonumber\\
&&\quad -i\int_{-\infty}^0 ds\; \left\{ \Sigma_{\hat{c}\hat{b}}(t-s)G_{\hat{b}a}(s)
+\Sigma_{\hat{c}\hat{c}}(t-s)G_{\hat{c}a}(s) \right\}. \nonumber
\end{eqnarray}
Now let us suppose that $F_{\hat{c}a}(t)$ can be taken to zero for some reason, then, by inserting
the expression for $G_{ac}(t)$ from 
Eq.~(\ref{gaa_gac}) into Eq.~(\ref{gaa_infty}), we obtain an equation for $G_{aa}(\infty)$,
which may have a nonvanishing solution signaling an ENE transition. 
This is essentially what happens in the field theory with linear FDR
as we will see in the next subsection. In the DM
field theory, however, the presence of the
first term on the right hand side of Eq.~(\ref{sd_cha}) spoils this scenario. 
In fact, as $t\to\infty$, $F_{\hat{c}a}(t)$ approaches
\[
G_{aa}(\infty)\int_0^\infty ds\; (-i)\Sigma_{\hat{c}a}(s),
\]
which is nonvanishing by assumption. Then the integral in Eq.~(\ref{gaa_infty}) becomes ill-defined
and we are forced to abandon the assumption of the non-zero $G_{aa}(\infty)$.
This finding is consistent with the recent nonpertubative proof by Das and Mazenko \cite{DM2} that the 
FNH will full spatial dependence does not support a sharp ENE transition. We note that the absence of the
ENE transition in our model is directly related to the presence of the self-energy 
$\Sigma_{\hat{c}a}$ in our model. This is also similar to the result of Ref.~\cite{DM2}, where
the self-energy coupling the hatted velocity field and the density field plays a crucial role
in removing the sharp transition.

\subsection{Field theory with linear FDR}

The SD equation in this case is given similarly to Eq.~(\ref{sd_general}) but now with the component
$\Phi$ taking eight different field variables, $\{\phi_i,\hat{\phi}_i\}$. So there are a lot more equations to consider in this approach.
The self-energies are, however, related to each other through many FDR, which can be obtained by 
applying the time reversal invariance to the SD equation. 
Here we list only the relevant FDR among the self-energies to our discussion. For
$\hat{\phi}=\hat{a},\hat{b},\hat{\eta}$ or $\hat{\theta}$, we have
\begin{eqnarray}
 &&\Sigma_{\hat{\phi}a}(t)=\frac{i}{T}\Theta(t)[A\Sigma_{\hat{\phi}\hat{a}}(t)
-\dot{\Sigma}_{\hat{\phi}\hat{\theta}}(t)] ,\nonumber \\
&&\Sigma_{\hat{\phi}b}(t)=\frac{i}{T}\Theta(t)[\frac{1}{a_0}\Sigma_{\hat{\phi}\hat{b}}(t)
-\dot{\Sigma}_{\hat{\phi}\hat{\eta}}(t)] ,\nonumber \\
&&\Sigma_{\hat{\phi}\eta}(t)=\frac{i}{T}\Theta(t)\Sigma_{\hat{\phi}\hat{b}}(t) ,\nonumber \\
&&\Sigma_{\hat{\phi}\theta}(t)=\frac{i}{T}\Theta(t)\Sigma_{\hat{\phi}\hat{a}}(t).
\label{fdr_fdr_2}
\end{eqnarray}

As in the DM approach, we study the time evolution of $G_{aa}(t)$ and its infinite-time limit
for the possible ENE transition. 
From the $(\Phi,\Phi^\prime)=(\hat{b}, a)$ component of the SD equation,
we have
\begin{eqnarray}
&&\dot{G}_{ba}(t)+\frac{\Gamma}{a_0} G_{ba} (t)-JAa_0 G_{aa}(t)-2i\Gamma T G_{\hat{b}a}(t) \nonumber \\
&&\quad\quad\quad\quad +\Gamma G_{\eta a}(t)-Ja_0 G_{\theta a}(t)=\widetilde{F}_{\hat{b}a}(t),
\label{Gaa:1}
\end{eqnarray}
where 
\begin{eqnarray}
\widetilde{F}_{\hat{b}a}(t)&=&-i\int_{-\infty}^{t} ds\; \sum_{\phi^\prime}
                     \Sigma_{\hat{b}\phi^\prime}(t-s)G_{\phi^\prime a}(s) \nonumber\\
                  &&-i\int_{-\infty}^{0} ds\;  \sum_{\hat{\phi}^\prime}
                     \Sigma_{\hat{b}\hat{\phi}^\prime}(t-s)
                      G_{\hat{\phi}^\prime a}(s).
\end{eqnarray}
Using the FDR in Eqs.~(\ref{fdr_fdr_1}) and (\ref{fdr_fdr_2}), 
we can rewrite the above quantity (multiplied by $T$) as
\begin{eqnarray}
T\widetilde{F}_{\hat{b}a}(t)&=&[\Sigma_{\hat{b}\hat{a}} \otimes (AG_{aa}+G_{\theta a})] (t)
                   \nonumber \\
                  &+&[\Sigma_{\hat{b}\hat{b}} \otimes (\frac 1 {a_0} G_{ba}+G_{\eta a})](t) 
                    \nonumber  \\ 
                 &-&[\Sigma_{\hat{b}\hat{\eta}} \otimes \dot{G}_{ba}](t) 
                 -[\Sigma_{\hat{b}\hat{\theta}} \otimes \dot{G}_{aa}](t), 
\label{f_bha_fdr}
\end{eqnarray}
where the convolution between two function $f(t)$ and $g(t)$ is defined by
\begin{equation}
[f \otimes g] (t) \equiv \int_0^{t} ds\; f(t-s) g(s).
\end{equation}
As in the DM approach, from the simple form of the equation for $a(t)$ and Eq.~(\ref{id}),
we can derive a nonperturbative relation, namely,
\begin{equation}
JG_{ba}(t)+\dot{G}_{aa}(t)=0.
\label{continuity_fdr}
\end{equation}

We now consider the $t\to\infty$ limit. 
If we assume as in the previous subsection that all the other correlation
functions except $G_{aa}(t)$ vanishes in this limit, then 
to the one-loop order, the only nonvanishing diagram in this limit
is again the one in Fig.~\ref{diagram}.
The contribution from this diagram to $\Sigma_{\hat{b}\hat{b}}(t)$
is $-2J^2A^2 G^2_{aa}(t)$. 
Taking the $t\to\infty$ limit in Eq.~(\ref{Gaa:1}) and using Eqs.~(\ref{f_bha_fdr})
and (\ref{continuity_fdr}), we obtain 
\begin{eqnarray}
-J A a_0 G_{aa}(\infty) &=& A \tilde{\sigma}G_{aa}(\infty) \nonumber \\
                   &-&   \frac{1}{JTa_0} \Sigma_{\hat{b}\hat{b}}(\infty)
                          \int_0^\infty ds\; \dot{G}_{aa}(s),
\end{eqnarray}
where 
\begin{equation}
\tilde{\sigma}=\frac{1}{T}\int_0^\infty ds\; \Sigma_{\hat{b}\hat{a}}(s).
\end{equation}
We therefore have
\begin{equation}
G_{aa}(\infty)=\frac{\Sigma_{\hat{b}\hat{b}}(\infty)}{J^2 \tilde{A} T a^2_0 }
               \left[ G_{aa}(\infty)-G_{aa}(0) \right],
\end{equation}
where $\tilde{A}=A(1+\tilde{\sigma}/(Ja_0))$.
Defining the nonergodicity parameter by $f=G_{aa}(\infty)/G_{aa}(0)$, we have
\begin{equation}
\frac{f}{1-f}=c_2 f^2,
\label{fo1mf}
\end{equation}
where
\begin{equation}
c_2=\frac{2A^2 [G_{aa}(0)]^2}{T\tilde{A}a^2_0}
\end{equation}
is a dimensionless quantity. This is exactly
the standard MCT equation for the nonergodicity parameter
in the so-called schematic model of the standard MCT \cite{Leuth,Bengtz}.
The nonergodic solution $f> 0$ exists when $c_2>4$.

The origin of the difference between the results of
the two field theoretic approaches on the existence of an ENE transition
can be traced back to the terms that are multiplied by the self-energy
$\Sigma_{\hat{b}\hat{b}}$. 
In both cases, this term is expressed as a time integral
of a correlation function.
In the field theory with linear FDR, this correlation function is proportional to a total time derivative
of $G_{aa}(t)$. From this, a well-defined equation like Eq.~(\ref{fo1mf}) follows for the nonergodicity parameter.
In the DM approach, however, the correlation function in the integrand is not a total time derivative of 
$G_{aa}(t)$, but contains an extra contribution from the self-energy coupling the hatted auxiliary field $\hat{c}$
and the density-like field $a$. On another level, we can understand that the difference comes from the fact
that, for the one-loop calculation in the field theory with 
linear FDR, only the first order terms are used
among those in the expression for the fields $\eta$ and $\theta$ in Eqs.~(\ref{eta_def})
and (\ref{theta_def}). This truncation of the non-polynomial action in the perturbation expansion 
does not occur in the DM field theory. This suggests that the cutoff of a sharp transition that the DM approach exhibits
at the one loop order is a kind of
nonpertubative information that only an infinite resummation in the field theory with linear FDR would have an access to.

\section{Numerical Calculations for the Das-Mazenko Field Theory}
\label{sec:numerical}

In order to study not just the nonergodicity parameter, but the full time evolution of $G_{aa}(t)$, 
it is desirable to have a single equation for the correlation function that accounts for the density feedback mechanism. 
In the original FNH in the DM field theoretic approach \cite{DM}, this was achieved only in the hydrodynamic limit.
It is, however, difficult even in the simple toy model to find a time evolution equation only for $G_{aa}(t)$.
In this section, we show that despite the lack of a complete set of linear FDR,
the DM field theoretical formulation presents a well-defined theory
at one-loop order. We do this by constructing a closed set of equations for
all the correlation functions involved and by solving them for $G_{aa}(t)$ numerically.
Because of the simple nature of the toy model, especially of
Eqs.~(\ref{np1}), (\ref{np2}), (\ref{fdr1}), (\ref{abc1}) and (\ref{abc2}), we only have five independent
correlation functions in the DM field theory. We choose them to be
$G_{aa}(t)$, $G_{a\hat{b}}(t)$, $G_{a\hat{c}}(t)$, $G_{c\hat{b}}(t)$
and $G_{c\hat{c}}(t)$. All the other correlation functions can be expressed in
terms of these five functions. 

If we define the Fourier transform by $\tilde{f}(\omega)=\int_{-\infty}^{\infty} dt \; e^{i\omega t} f(t)$
for an arbitrary function $f(t)$, the Fourier transforms of these five correlation functions can be written as
\begin{eqnarray}
 &&\widetilde{G}_{aa}(\omega)=\frac{J^2}{|D(\omega)|^2}
[|a_R(\omega)|^2(2\Gamma T-\widetilde{\Sigma}_{\hat{b}\hat{b}}(\omega))
\label{gaa_o}\\
&& \quad\quad\quad
-2\; \mathrm{Re}\{a_R(\omega) \Gamma^\ast_R(\omega) \widetilde{\Sigma}_{\hat{b}\hat{c}}(\omega)\}
-|\Gamma_R(\omega)|^2\widetilde{\Sigma}_{\hat{c}\hat{c}}(\omega)],
\nonumber  \\
&&\widetilde{G}_{a\hat{b}}(\omega)=\frac{-iJa_R(\omega)}{D(\omega)},
\label{gabh_o}\\
&&\widetilde{G}_{a\hat{c}}(\omega)=\frac{-iJ\Gamma_R(\omega)}{D(\omega)},
\label{gach_o}\\
&&\widetilde{G}_{c\hat{b}}(\omega)=\frac{\omega+J\widetilde{\Sigma}_{\hat{c}a}(\omega)}{D(\omega)},
\label{gcbh_o}\\
&&\widetilde{G}_{c\hat{c}}(\omega)=\frac{i\omega^2-iJK_R(\omega)}{D(\omega)},
\label{gcch_o}
\end{eqnarray}
where $\mathrm{Re}$ denotes the real part and
\begin{eqnarray}
&&a_R(\omega)=a_0-i\widetilde{\Sigma}_{\hat{c}c}(\omega) ,
\label{a_R}\\
&&\Gamma_R(\omega)=\Gamma+i\widetilde{\Sigma}_{\hat{b}c}(\omega) ,\\
&&K_R(\omega)=JAa_0-i\widetilde{\Sigma}_{\hat{b}a}(\omega) ,\\
&&D(\omega)=a_R(\omega)(\omega^2-JK_R(\omega)) \nonumber \\
&&\quad\quad\quad+i\Gamma_R(\omega)
(\omega+i\widetilde{\Sigma}_{\hat{c}a}(\omega)).
\label{Do}
\end{eqnarray}
Note that the Fourier transforms of seven self-energies
are involved in the above equations, which are
$\widetilde{\Sigma}_{\hat{b}a}(\omega)$,
$\widetilde{\Sigma}_{\hat{b}c}(\omega)$,
$\widetilde{\Sigma}_{\hat{c}a}(\omega)$,
$\widetilde{\Sigma}_{\hat{c}c}(\omega)$,
$\widetilde{\Sigma}_{\hat{b}\hat{b}}(\omega)$, 
$\widetilde{\Sigma}_{\hat{b}\hat{c}}(\omega)$ and
$\widetilde{\Sigma}_{\hat{c}\hat{c}}(\omega)$. 
At the one-loop order of the perturbation theory, these seven self-energies are given by
functions of the five independent correlation functions thus yielding a closed
set of equations. It is more convenient to present the one-loop self-energies in
the time domain. 
(The detailed derivation of the one-loop self-energies in terms of the correlations functions
will be given elsewhere \cite{yeo2}.) 
They are given by
\begin{eqnarray}
\Sigma_{\hat{b}a}(t)&=&-2i a_0 J^2 A T G_{a\hat{b}}(t)G_{c\hat{b}}(t)
                +i a_0 J TG_{c\hat{b}}(t) G_{c\hat{c}}(t)\nonumber \\
&-&4J^2 A^2 G_{aa}(t) G_{a\hat{b}}(t)
-2i J A T G_{a\hat{b}}(t)G_{a\hat{c}}(t), \nonumber \\
&+&\frac{1}{2}JTG_{c\hat{b}}(0^+)\delta(t),
\label{self1} 
\end{eqnarray}
\begin{eqnarray}
&&\Sigma_{\hat{b}c}(t)=-ia^2_0 J^2 T \left(G_{c\hat{b}}(t)\right)^2
                                      -ia_0 J T G_{a\hat{b}}(t)G_{c\hat{c}}(t) \nonumber\\
	&& \quad\quad +2i a_0 J^2 A T \left(G_{a\hat{b}}(t)\right)^2 
             -2 J A G_{aa}(t)G_{a\hat{c}}(t), 
\label{self2}
\end{eqnarray}
\begin{eqnarray}
\Sigma_{\hat{c}a}(t)&=&-2 J A G_{aa}(t)G_{c\hat{b}}(t)
          +2i JAT \left(G_{a\hat{b}}(t)\right)^2 \nonumber \\
	&-&i T G_{a\hat{b}}(t)G_{c\hat{c}}(t) 
	 -i T G_{a\hat{c}}(t)G_{c\hat{b}}(t),
\label{self3}
\end{eqnarray}
\begin{eqnarray}
\Sigma_{\hat{c}c}(t)&=&2i a_0 J T G_{a\hat{b}}(t)G_{c\hat{b}}(t)
                           +iT G_{a\hat{b}}(t)G_{a\hat{c}}(t) \nonumber \\
                          &-&   G_{aa}(t)G_{c\hat{c}}(t),
\label{self4}
\end{eqnarray}
\begin{eqnarray}
\Sigma_{\hat{b}\hat{b}}(t)&=&-2J^2 A^2  \left(G_{aa}(t)\right)^2
                           - 2 a_0 J^2 A T^2 \left(G_{a\hat{b}}(t)\right)^2 \nonumber \\
			&+& \frac{1}{2} a^2_0 J^2 T^2 \left(G_{c\hat{b}}(t)\right)^2,
\label{self5}
\end{eqnarray}
\begin{equation}
\Sigma_{\hat{b}\hat{c}}(t)=-2i JAT  G_{aa}(t)G_{a\hat{b}}(t)
                           +a_0 J T^2 G_{a\hat{b}}(t)G_{c\hat{b}}(t),
\label{self6}
\end{equation}
\begin{equation}
\Sigma_{\hat{c}\hat{c}}(t)=- iT G_{aa}(t)G_{c\hat{b}}(t)
                           -T^2 \left(G_{a\hat{b}}(t)\right)^2.
\label{self7}
\end{equation}
Note that these expressions are valid only for $t>0$. The self-energies in Eqs.~(\ref{self1})-(\ref{self4}) vanish for
$t<0$ due to causality. The remaining self-energies satisfy
$\Sigma_{\hat{b}\hat{b}}(-t)=\Sigma_{\hat{b}\hat{b}}(t)$, 
$\Sigma_{\hat{c}\hat{c}}(-t)=\Sigma_{\hat{c}\hat{c}}(t)$, and
$\Sigma_{\hat{b}\hat{c}}(-t)=-\Sigma_{\hat{b}\hat{c}}(t)$.

\begin{figure}
 \begin{center}
\includegraphics[width=0.3\textwidth]{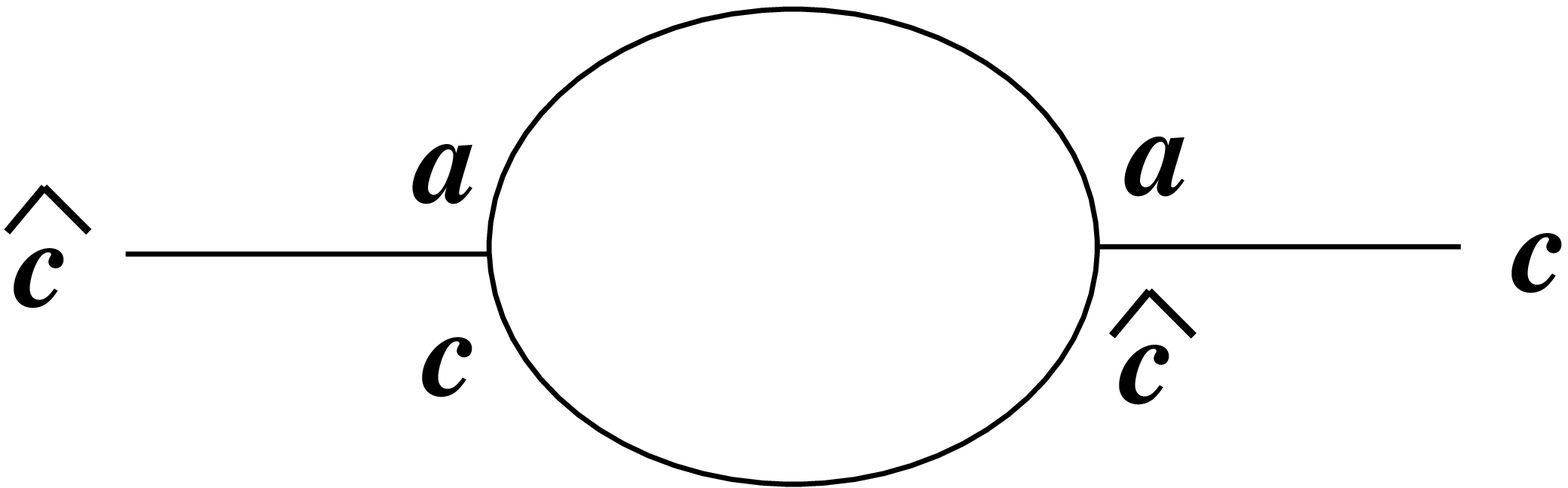}
\end{center}
\caption{The one-loop diagram contributing to $\widetilde{\Sigma}^{\prime\prime}_{\hat{c}c}(\omega)$
in the $\omega\to\infty$ limit}
\label{diagram1}
\end{figure}

There is one point that requires caution in performing a numerical calculation on these types of self-consistent equations. 
We note that $\widetilde{G}_{c\hat{c}}(\omega)$ does not decay to zero as $\omega\to \infty$. This 
suggests that there is a delta-function singularity in $G_{c\hat{c}}(t)$ at short time, which has to be
treated separately in a numerical calculation. 
We write $G_{c\hat{c}}(t)=i\alpha\delta(t)+$ regular terms for some real constant $\alpha$.
If we denote by $f^\prime$ and $f^{\prime\prime}$ 
the real and imaginary parts of a complex function $f$, respectively, then we can write
\begin{equation}
\widetilde{G}^{\prime\prime}_{c\hat{c}}(\omega)=
\alpha+\widetilde{G}^{\prime\prime(\mathrm{reg})}_{c\hat{c}}(\omega),
\end{equation}
where 
$\lim_{\omega\to\infty}\widetilde{G}^{\prime\prime(\mathrm{reg})}_{c\hat{c}}(\omega) =0$.
From Eqs.~(\ref{gcch_o}), (\ref{a_R}) and (\ref{Do}), we find that 
$\alpha^{-1}=a_0+\lim_{\omega\to\infty} \widetilde{\Sigma}^{\prime\prime}_{c\hat{c}}(\omega)$.
At the one-loop order, a nonvanishing contribution to $\widetilde{\Sigma}^{\prime\prime}_{c\hat{c}}(\omega)$
in the infinite-$\omega$ limit comes from
the diagram depicted in Fig.~\ref{diagram1}. We thus have
\begin{equation}
\lim_{\omega\to\infty} \widetilde{\Sigma}^{\prime\prime}_{c\hat{c}}(\omega)
= -\alpha\int_{-\infty}^{\infty}\frac{d\omega}{2\pi} \widetilde{G}_{aa}(\omega)
=-\alpha G_{aa}(0),
\end{equation}
and
\begin{equation}
 \frac{G_{aa}(0)}{a^2_0}=\frac{1}{a_0\alpha}-\frac{1}{(a_0\alpha)^2}
\label{gaa0}
\end{equation}
is the initial value of the dimensionless correlation function $G_{aa}(t)/a^2_0$.
Therefore, the initial value of the correlation function is determined self-consistently in the present
field theoretic approach. This is in contrast to 
other approaches \cite{KK,NH} where the static limit of the correlation functions was used as an input
for the initial condition.
Note that $G_{aa}(0)/a^2_0\leq 1/4$ where the maximum value
occurs when $a_0\alpha=2$. We believe that this peculiar behavior is due to the one-loop perturbation theory and that
if we consider a higher-loop theory this condition will certainly change.  

We now present numerical solutions to the above coupled equations. 
We start by evaluating the one-loop self-energies using Eqs.~(\ref{self1})-(\ref{self7}) from
some appropriate initial form of the correlation functions (e.\ g.\ the bare correlation functions)
given as functions of time.
We make Fourier transforms of these self-energies and update 
the correlation functions by using Eqs.~(\ref{gaa_o})-(\ref{gcch_o}). Then, the inverse Fourier transforms
are performed to compare the input and output correlation functions. This procedure is repeated until
the convergence is achieved. We find that the convergence is achieved in less than 100 iterations
in most cases. The number $N$ of mesh points used in the time and frequency integrals ranges
from 8000 to 26000. The cutoffs, $\Lambda_t$ and $\Lambda_\omega$, for the time
and frequency integrals, respectively, must be adjusted so that all the five correlation functions
and the seven self-energies are accommodated both in the time and frequency spaces. 
We maintain that
$\Lambda_t \Lambda_\omega \sim N \pi$ to have a consistent numerical Fourier transform \cite{Lado}.  
As we will explain below, $G_{aa}(t)$ can in general be a relatively long-ranged function, but other correlation functions
such as $G_{c\hat{b}}(t)$ and $G^{\rm (reg)}_{c\hat{c}}(t)$ are short-ranged so that
we need large $\Lambda_\omega$ for those functions. As functions
of short and long-ranged are mixed in the calculations, both $\Lambda_t$ and
$\Lambda_\omega$ must be sufficiently large.

For a numerical calculation of the above coupled set of equations, we need to put everything
in dimensionless forms. From Eqs.~(\ref{eq:a}) and (\ref{eq:b}), we see that
$a_0/\Gamma$ has the dimension of time. Once we put all the correlation and response functions
in their respective dimensionless forms, we find that the self-consistent equations are completely described by 
two dimensionless parameters, $\kappa$ and $\tilde{T}$
defined by
\begin{equation}
\kappa\equiv\left( \frac{Ja_0}{\Gamma}\right)^2 a_0 A, \quad\quad\quad
\tilde{T}\equiv\left( \frac{J}{\Gamma}\right)^2 a_0 T .
\end{equation}

\begin{figure}
 \begin{center}
\includegraphics[width=0.45\textwidth]{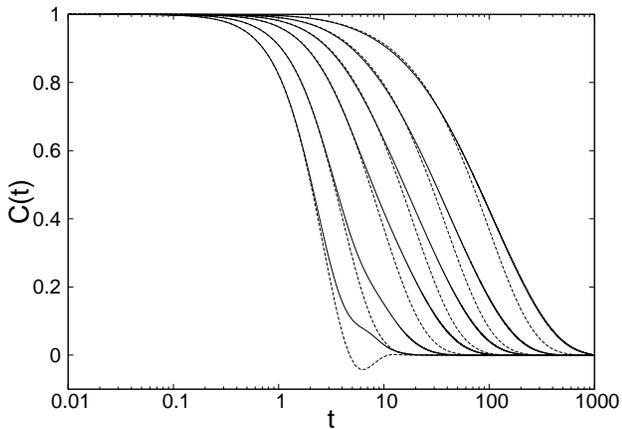}
\caption{Normalized correlation function $C(t)$ as a function of time 
$t$ measured in units of $a_0/\Gamma$ for fixed $\widetilde{T}/\kappa=0.1$. 
The solid lines are the solutions to the self-consistent one-loop equations for 
$\widetilde{T}=0.05, 0.025, 0.01, 0.005, 0.0025$
and 0.001 from left to right. The dashed lines are the corresponding bare correlation functions.}
\label{tk=0.1}
\end{center}
\end{figure}

\begin{figure}
 \begin{center}
\includegraphics[width=0.45\textwidth]{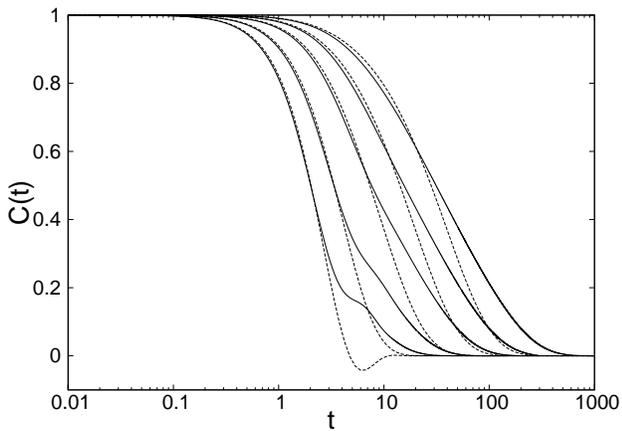}
\caption{
Normalized correlation function $C(t)$ as a function of time 
$t$ measured in units of $a_0/\Gamma$ for fixed $\widetilde{T}/\kappa=0.2$. 
The solid lines are the solutions to the self-consistent one-loop equations for 
$\widetilde{T}=0.1, 0.05, 0.02, 0.01$
and 0.005 from left to right. The dashed lines are the corresponding bare correlation functions.}
\label{tk=0.2}
\end{center}
\end{figure}

In Figs.~\ref{tk=0.1} and \ref{tk=0.2}, we plot the normalized correlation function
$C(t)\equiv G_{aa}(t)/G_{aa}(0)$ which corresponds to the density auto-correlation function in the FNH
for various values of the parameters $\kappa$ and $\widetilde{T}$.
These are compared with the corresponding bare correlation functions. 
From these figures, we can see that in general the one-loop correlation functions
are more stretched in later times compared to the bare correlation functions. 
We perform numerical calculations for fixed value of $\widetilde{T}/\kappa=T/(a^2_0 A)$. 
The analytic expressions for the bare correlation
functions can easily be obtained from the Gaussian part of
Eq.~(\ref{S_DM}), and we note that
the initial value of the bare function $G^{(0)}_{aa}(t)$ is given by 
$G^{(0)}_{aa}(0)/a^2_0 =  \widetilde{T}/\kappa$ in dimensionless quantities. 
As we can see by comparing Figs.~\ref{tk=0.1} and \ref{tk=0.2},
the difference between the renormalized correlation functions and the bare ones is small
for small $\widetilde{T}/\kappa$, but the nonlinear effects increase with increasing $\widetilde{T}/\kappa$.
The value of $a_0\alpha$ in Eq.~(\ref{gaa0}) is also determined from the numerical calculation. For
fixed $\widetilde{T}/\kappa$, we find that $a_0\alpha$ and thus $G_{aa}(0)/a^2_0$ is almost constant
when we change $\widetilde{T}$. As we increase $\widetilde{T}/\kappa$, the initial value
$G_{aa}(0)/a^2_0$ increases as well.
We find that the numerical solutions for the self-consistent equations exist
only for $\widetilde{T}/\kappa$ less than some maximum value which is found to be around 0.37
for the one-loop theory when $G_{aa}(0)/a^2_0$ reaches it maximum value $1/4$. 
We note that this has nothing to do with the ENE singularity which we discussed earlier, since
it involves the short-time behavior of the correlation functions.
Indeed, as we approach this maximum value, the response functions, 
$G_{c\hat{b}}(t)$ and $G^{\rm (reg)}_{c\hat{c}}(t)$ become increasingly short-ranged in time.
For example, the initial time derivative $\dot{G}_{c\hat{b}}(0^+)$ approaches $-\infty$ as 
$\widetilde{T}/\kappa$ approaches 0.37. 
We can understand these behaviors by investigating carefully the $t\to 0^+$ limit of the SD
equations (\ref{sd_general}) \cite{yeo2}. 
We believe that the particular values of the parameters
are specific to the one-loop calculation and will change as higher-loop contributions are considered.

As expected from the discussion in the previous section, 
$C(t)$ shows a completely ergodic behavior decaying to zero
as $t\to\infty$ for all parameter values in our numerical calculations.
We note that,
in the standard MCT \cite{Gotze0,KM,Das,RC} above the ENE transition, the density auto-correlation
function exhibits a plateau before its eventual decay in time. This is not obvious in our numerical
results. We believe that the present toy model, without having a realistic
wavenumber dependence, is too simple to capture the plateau, if any, 
in the density auto-correlation function calculated from
the one-loop approximation. In order to see if the DM 
field theoretical approach to FNH produces a plateau in a given 
order of the loop expansion, one would have to solve wavenumber-dependent 
versions of Eqs. (\ref{gaa_o})-(\ref{self7}). This remains to be seen in the future study.

\section{Discussion and Conclusion}

In this paper, we have introduced and studied the toy model of FNH of supercooled liquids containing only a couple of dynamical
field variables $a(t)=a_0+\delta a(t)$ and $b(t)$ without any spatial dependence. We have developed
two different field theoretic formulations of this model first by following the DM prescription and then the 
method involving a set linear FDR. We were also able to perform numerical calculations on the coupled 
equations for the correlation and response functions in the DM field theory at the one-loop order. 

The major difference between the two field theoretic formulations is
the way of treating the density nonlinearities that appears in the problem in the form of $1/a(t)$. 
In the DM field theory this is treated in a simple way by introducing a single
auxiliary field $c(t)$ and its hatted counterpart $\hat{c}(t)$. 
On the other hand, in the field theory with linear FDR, a couple
of auxiliary fields $\eta(t)$ and $\theta(t)$, as well as their hatted partners, are introduced in such a way
that the dynamical action becomes invariant under a set of time-reversal transformations. 
This manipulation results in the dynamical action with terms which are 
non-polynomial functions of the main field variables $\delta a(t)$ and $b(t)$.
We note that these terms are 
proportional to $\hat{\eta}(t)$ and $\hat{\theta}(t)$ fields.
Therefore, in the renormalized 
perturbation theory at a given order of the loop expansion, one has to truncate
the non-polynomial functions at the appropriate order. 
In some sense, we might say that, because of the truncation, the effect of the density nonlinearities 
would not be fully incorporated into the field theory at any finite 
order of the perturbation theory. We believe that this is one of the reasons why
the sharp ENE-type transition appears at the one-loop order in this formulation. 

This is in contrast to the DM field theory where no truncation of the dynamical
action is necessary when the loop expansion is performed. 
The loop expansion, therefore, has a different meaning from the field theory with linear FDR.
We have shown that 
the self-energy $\Sigma_{\hat{c}a}(t)$ which
corresponds to the one that couples the density and the hatted velocity fields
plays a key role in removing the sharp transition in the one-loop calculation. This is 
consistent with the original finding by DM \cite{DM} and with 
the recent nonperturbative analysis of the FNH \cite{DM2}.
This fact has often been interpreted as the coupling between the current and the density being responsible
for the ergodicity restoring mechanism. 
In our toy model, the field $\hat{c}(t)$
is introduced in the DM field theory to take care of the density nonlinearities 
as $\hat{\eta}(t)$ and $\hat{\theta}(t)$
are in the other formulation. Even though we only perform the one-loop
calculation, we can say that the self-energy $\Sigma_{\hat{c}a}(t)$ contains
the nonperturbative information arising from the density nonlinearities which are not present in
the field theory with linear FDR. In this respect, it may be more appropriate to regard the cutoff mechanism
resulting from the full nonperturbative treatment of the density nonlinearities 
than from the coupling between the current and the density. The recent numerical
calculation \cite{Gupta} where a direct integration of the generalized Langevin equations in FNH
are performed also demonstrates that the $1/\rho$-nonlinearities ($\rho$ is the density) are playing an 
essential role restoring the ergodic behavior in supercooled liquids. 

It is sometimes discussed in literatures that the DM field theory is inconsistent with
the FDR. However, a set of FDR does hold in the DM field theory, 
which is given in Eq. (\ref{fdr1}) for our toy model. In the field theory with linear FDR,
a larger set of FDR exists as in Eqs. (\ref{fdr_fdr_1}) and (\ref{fdr_fdr_2}). 
In the original analysis of 1986 \cite{DM}, Das and Mazenko used another FDR 
(linking $G_{aa}$ and $G_{a\hat{a}}$ in the toy model notation)
in addition to Eq. (\ref{fdr1}) and simplified the equations involved. 
This is valid only in the hydrodynamic limit. 
In the present paper, we do not use such 
additional simplifications. Instead, we keep only Eq. (\ref{fdr1}), and take the equations
which the correlation functions satisfy, Eqs. (\ref{gaa_o})-(\ref{self7}), as a set of
self-consistent equations, and solve them numerically in Sec.~\ref{sec:numerical} .
This program could be generalized to a realistic situation where
the full spatial dependence is present.
What we have shown in this paper is that the DM field theory, viewed 
as a collection of self-consistent equations for the correlation functions
at a given order of loop expansion, is a well-defined field theoretic approach to glass forming
liquids.

There are other field theoretic approaches to the FNH than those considered in this paper. 
A similar toy model of the FNH to ours but
of different nature has been studied by Kim and Kawasaki \cite{KKtoy}
some time ago. In this model, the $N$-component of density-like and the
$M$-component momentum-like field variables are
introduced without spatial dependence. They consider the limit where the numbers $N$ and $M$ 
approach infinity, and find that a sharp transition is present when the condition $M<N$ is maintained
in the limiting process, while it is absent when $M=N$.  
In Ref.~\cite{SDD},
a simpler version of the FNH than that of DM was considered, where the sharp transition was found to be absent.
We note, however, that in both cases
the effective free energy is quadratic in both density and the momentum variables from the outset. 
There is no need to introduce the auxiliary fields and a full set of linear FDR exist in these models. 
Therefore the kind of density nonlinearities discussed in this paper is not present and
the absence of the sharp transition found in these works is probably of a different origin.

The work by Mayer et al.~\cite{Mayer}
is another interesting zero-dimensional model for glass forming liquids.
It is in general hard to make a direct connection between the projection
operator approach, in which Ref.~\cite{Mayer} is set, and the field
theoretical one. It is, however, clear from Ref.~\cite{Mayer} that a 
nonperturbative effect that comes from considering an infinite number 
of equations is responsible for cutting off the sharp transition. 
In Ref.~\cite{Mayer}, the sharp transition is always present when one considers only
a finite number of equations. This is related to our finding that, 
when the density nonlinearities are treated in the field theory with 
linear FDR within the loop expansion, one has to truncate a non-polynomial 
function, and the sharp transition follows. 
We might say that the nonperturbative information that 
cuts off the sharp transition is somehow preserved in the DM approach,
since one can avoid truncating the dynamical action at a given order
of the loop expansion. 

There are many ways in which the present result can be generalized. 
An obvious generalization is to consider a higher-order
perturbation theory. Because of the simple nature of the model, one can without much difficulty
construct the higher-loop DM field theory of the model and perform the numerical calculations
as done in this paper.
It will be interesting to see how the one-loop results, especially the particular initial values of the
correlation functions, get changed when the higher-loop contributions are considered.
We believe that the numerical methods developed in Sec.~\ref{sec:numerical} for treating 
the coupled equations for the correlation and response functions, especially those 
concerning the short-time behavior of functions, will prove to be useful for 
an eventual application to the full wavenumber dependent FNH in the future.




\begin{acknowledgments}
We would like to thank Bongsoo Kim and Shankar Das for useful discussions.
This work was supported by the Korea Research Foundation Grant funded by the Korean
Government (MOEHRD, Basic Research Promotion Fund) (KRF-2008-313-C00333).
\end{acknowledgments}



\end{document}